\documentclass{moriond}
\usepackage{siunitx}
\usepackage{units}
\usepackage{upgreek}
\usepackage{tabularx}
\usepackage{hyperref}

\def\be{\begin{equation}}
\def\ee{\end{equation}}
\def\bea{\begin{eqnarray}}
\def\eea{\end{eqnarray}}

\usepackage{xspace}

\def\vB{\vec{v} \times \vec{B}}

\def\xmax{X$_\mathrm{max}$\xspace}

\newcommand{\Photo}{}

\begin{document}
\vspace*{4cm}
\title{{A novel method for the absolute energy calibration\\of large-scale cosmic-ray detectors\\using radio emission of extensive air showers}}

\author{Christian Glaser$^\mathrm{1,}$\footnote{ \href{mailto:glaser@physik.rwth-aachen.de}{glaser@physik.rwth-aachen.de}} for the Pierre Auger Collaboration$^{2, }$\footnote{Full author list is available at \url{http://auger.org/archive/authors_2017_03.html}}}

\address{$^1$RWTH Aachen University, Aachen, Germany\\
$^2$Observatorio Pierre Auger, Av. San Martin Norte 304, 5613 Malarg\"ue, Argentina\\
\quad}

\maketitle\abstracts{
Ultra-high energy cosmic rays impinging onto the atmosphere induce huge cascades of secondary particles. The measurement of the energy radiated by these air showers in form of radio waves enables an accurate measurement of the cosmic-ray energy. Compared to the well-established fluorescence technique, the radio measurements are less dependent on atmospheric conditions and thus potentially reduce the systematic uncertainty in the cosmic-ray energy measurement significantly. Two attractive aspects are that the atmosphere is transparent to MHz radio waves and the radio emission can be calculated from first-principles using classical electrodynamics. \newline
This method will be discussed for the Engineering Radio Array (AERA) of the Pierre Auger Cosmic-Ray Observatory. AERA detects radio emission from extensive air showers with energies beyond \SI{e17}{eV} in the 30 - 80 MHz frequency band and consists of more than 150 autonomous radio stations covering an area of about \SI{17}{km^2}. It is located at the same site as the Auger low-energy detector extensions enabling combinations with various other measurement techniques. %
}

\section{Introduction}
Ultra-high energy cosmic rays (UHECRs) impinging onto the atmosphere induce huge cascades of secondary particles. Established techniques for their detection are the measurement of the particles of the air shower that reach the ground or the observation of the isotropic fluorescence light emitted by molecules that have been excited by the shower particles \cite{Auger2014,TA2008}. An important observable for most analyses of high-energy cosmic rays is their energy and in particular the accuracy, i.e. the systematic uncertainty, of the energy measurement.
So far, the Pierre Auger Observatory has used the fluorescence technique for the absolute energy calibration. Telescopes measure the fluorescence light emitted by air showers which is proportional to the calorimetric shower energy and allows for an accurate determination of the energy of the primary particle with a systematic uncertainty of 14\% \cite{EnergyScaleICRC2013}. However, fluorescence light detection is only possible at sites with good atmospheric conditions, and precise quantification of scattering and absorption of fluorescence light under changing atmospheric conditions requires extensive atmospheric monitoring efforts \cite{Auger2014}.

Another independent method for the detection of cosmic rays is the detection of broadband radio emission from air showers which is an emerging field of research \cite{Huege2016,Schroeder2016a}. The radio technique combines a duty cycle close to 100\% with an accurate and precise measurement of the cosmic-ray energy \cite{AERAEnergyPRL,AERAEnergyPRD} as well as a good sensitivity to the mass of the primary cosmic-ray \cite{LOFARNature2016}. In particular, the energy measurement outperforms existing techniques in terms of achievable accuracy and can reduce the systematic uncertainty to 10\% \cite{PhDGlaser}. This is mostly due to the lack of absorption of radio waves in the atmosphere and the corresponding insensitivity to changing environmental conditions, and because the radio emission can be calculated theoretically via first principles from the air-shower development. In the following, we will discuss how the radio technique can be used for the absolute energy calibration and discuss its systematic uncertainties.

\section{Radio emission from extensive air showers}
\label{sec:radioemissionmechanisms}
Two processes of the radio emission of air showers have been identified. The dominant \emph{geomagnetic} emission arises from the deflection of electrons and positrons in the shower front at the Earth's magnetic field and is polarized along the direction of the Lorentz force ($\propto\, \vec{v} \times \vec{B}$) that is acting on the charged particles \cite{KahnLerche1966}. The field strength of the emission scales with the absolute value of the geomagnetic field $\vec{B}$ and the sine of the angle $\upalpha$ between the shower direction $\vec{v}$ and the geomagnetic field. Muons are also deflected in principle, but due to their much lower charge/mass ratio they do not contribute significantly to the radio emission \cite{Huege2016}. 

\begin{figure}[t]
\centering
\includegraphics[width=0.65\textwidth]{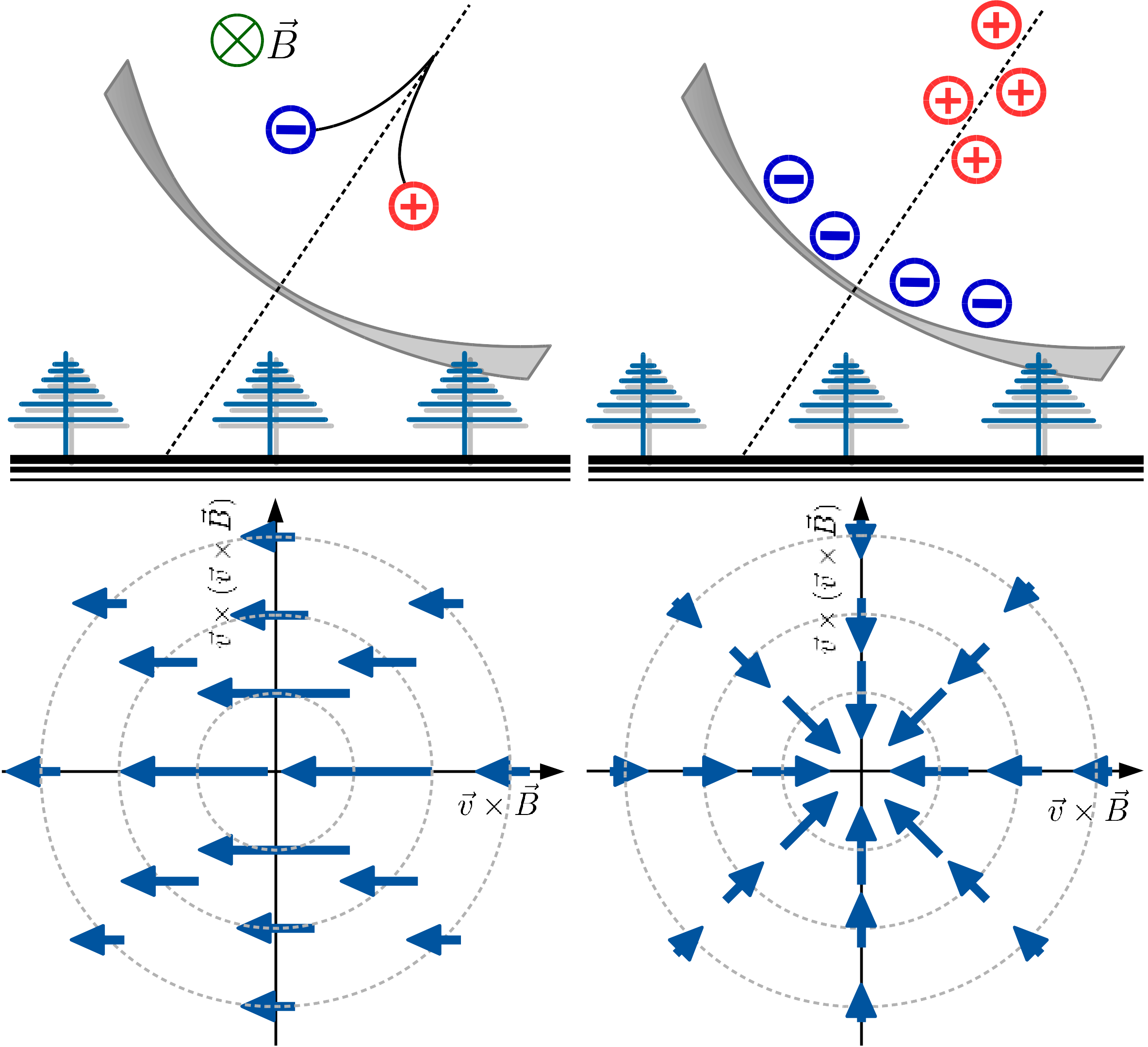}
\caption{Sketch of the two mechanisms of radio emission from air showers. (left) The dominant geomagnetic emission and (right) the sub-dominant charge-excess emission. The lower plot shows the polarization of the radio pulses originating from the two processes. The polarization is shown in a coordinate system in the shower plane where one axis is aligned in the $\vB$ direction and the other axis perpendicular to it ($\vec{v}$ is the shower direction and $\vec{B}$ the geomagnetic field).}
\label{fig:emission_mechanisms}
\end{figure}

The second, subdominant emission arises from a time-varying negative charge-excess in the shower front which is polarized radially with respect to the axis of the air shower and is referred to as the \emph{charge-excess} or \emph{Askaryan} effect \cite{Askaryan1962}. An illustration of the two emission mechanisms is given in Fig.~\ref{fig:emission_mechanisms}. The relative strength of the charge-excess emission  varies with the absolute value of the geomagnetic field, the incoming direction of the air shower and the distance from an observer to the shower axis and can take values ranging from a few percent up to almost 50\% \cite{GlaserErad2016}. The charge-excess strength can be  measured with the polarization of the signal pulse and was determined to have an average value of $14\% \pm 2\%$ at the Pierre Auger Observatory \cite{AERAPolarization}.

In addition to the two emission processes, the refractive index of air has a significant impact on the radio emission.
As the refractive index of air is larger than unity, signals emitted at different stages of the shower development arrive at the same time at observers that see the shower under a specific angle, the so-called \emph{Cherenkov angle}. This leads to an increased signal strength on a ring around the shower axis, the \emph{Cherenkov ring}, and leads to coherent emission up to \si{GHz} frequencies. As the refractivity is not constant but scales with the air density towards higher altitudes, the ring is smeared out. 
Due to coherence effects, the radio emission is strongest below \SI{100}{MHz}. At larger frequencies the emission is less coherent resulting in smaller signal strengths. 
Although the radio signal is not Cherenkov light and emitted by different mechanisms, at frequencies well beyond \SI{100}{MHz} the emission can be detected only in very specific geometries where observers are at the Cherenkov angle \cite{LOFARCherenkov2015}. In addition, below \unit[30]{MHz} atmospheric noise and short-wave band transmitters make measurements unfeasible and above \unit[80]{MHz} the FM band interferes. Hence, in most experimental setups (e.g. AERA), the ``golden'' frequency band between \unit[30 - 80]{MHz} is used \cite{AERAEnergyPRD}.

\begin{figure}[t]
\centering
\includegraphics[width=0.55\textwidth]{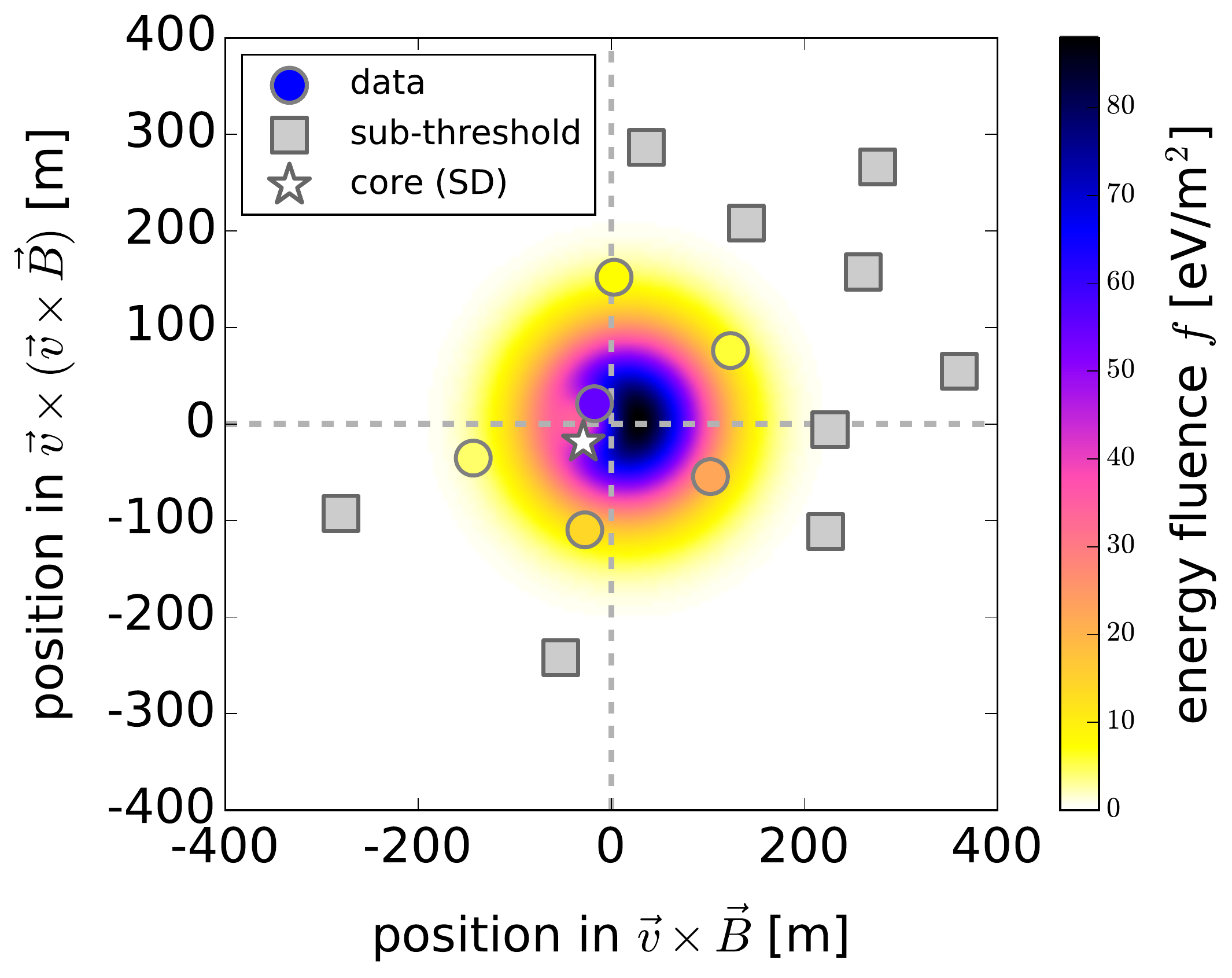}
\includegraphics[width=0.44\textwidth]{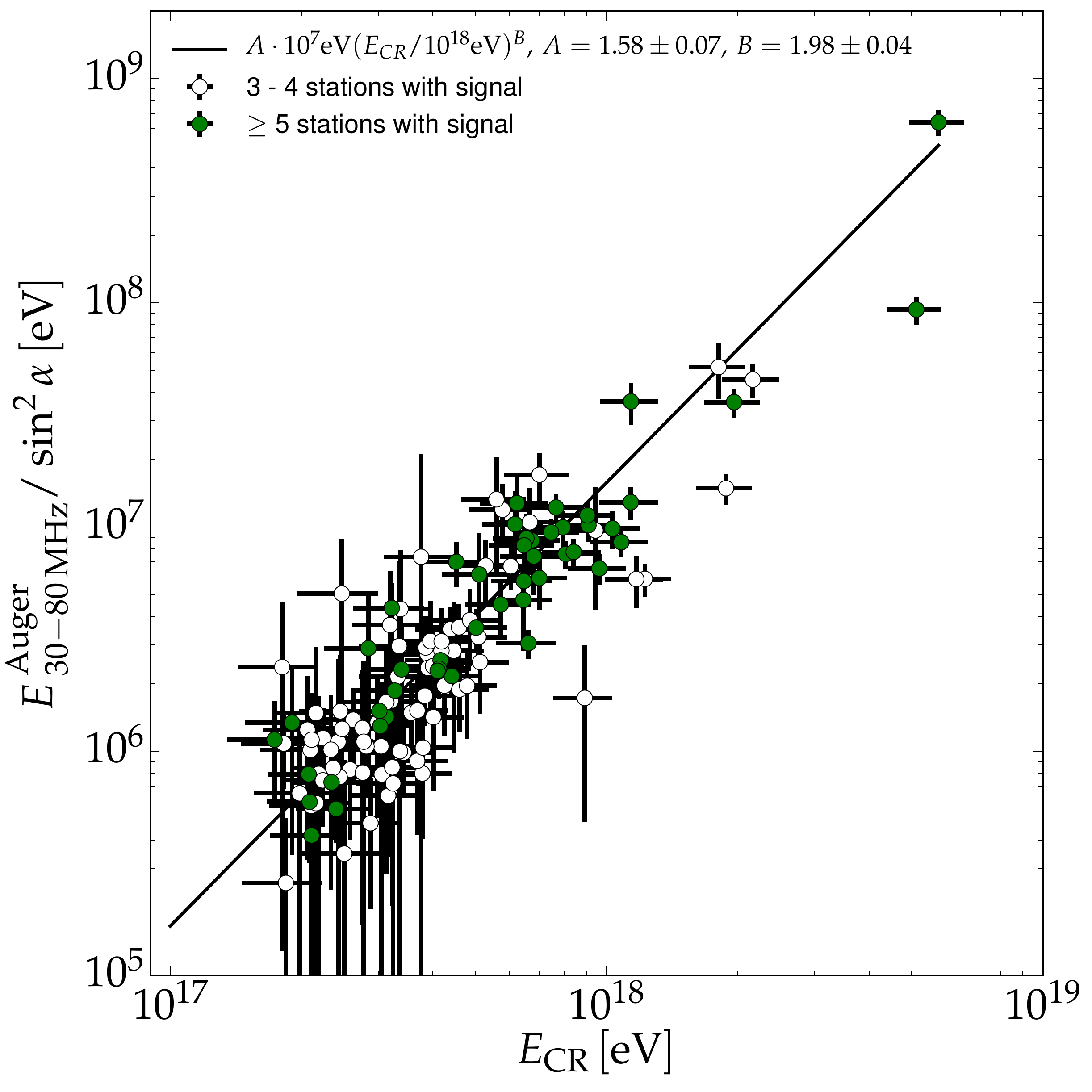}
\caption{(left) Energy fluence for an extensive air shower with an energy of \unit[$4.4 \times 10^{17}$]{eV}, and a zenith angle of 25$^{\circ}$ as measured in individual AERA radio detectors (circles filled with color corresponding to the measured value) and fitted with the azimuthally asymmetric, two-dimensional signal distribution function (background color). Both, radio detectors with a detected signal (\emph{data}) and below detection threshold (\emph{sub-threshold}) participate in the fit. The best-fitting impact point of the air shower is at the origin of the plot, slightly offset from the one reconstructed with the Auger surface detector (\emph{core (SD)}). 
(right) Correlation between the normalized radiation energy and the cosmic-ray energy $E_{\text{CR}}$ as determined by the Auger surface detector. Open circles represent air showers with radio signals detected in three or four radio detectors. Filled circles denote showers with five or more detected radio signals. Figures and captions from  \protect{\cite{AERAEnergyPRL}}.}
\label{fig:LDF}
\end{figure}

As a consequence of the superposition of the two emission mechanisms, the lateral distribution function (LDF) of the electric-field strength has been found to have a radial asymmetry \cite{LOFARLDF}. The two-dimensional shape of the LDF is best understood in a coordinate system with one axis perpendicular to the shower direction $\vec{v}$ and Earth's magnetic field $\vec{B}$ (along the Lorentz force $\sim \vec{v} \times \vec{B}$), and the perpendicular axis $\vec{v}\times (\vec{v}\times \vec{B})$. In this coordinate system, as visible from the sketch of Fig.~\ref{fig:emission_mechanisms}, the two emission mechanisms interfere destructively left of the shower axis and interfere constructively right of the shower axis. An example of this asymmetric signal distribution is shown in Fig.~\ref{fig:LDF}. This peanut-like shape can be modeled analytically with an empirical function consisting of two two-dimensional Gaussians \cite{LOFARLDF}. The function has nine free parameters. However, five of them can be fixed using predictions from Monte-Carlo simulations which results in a large practical usability of this parametrization\cite{LOFAREnergy,AERAEnergyPRD}, e.g., the amplitude or the integral of the function can be used to determine the cosmic-ray energy and the width of the function depends on the distance to the shower maximum \xmax.

\subsection{Calculation of radiation energy release}
\label{sec:theoryerad}

The total amount of radio emission emitted by an air shower can be quantified using the concept of radiation energy \cite{AERAEnergyPRL,AERAEnergyPRD} which is the energy emitted by the air shower in the form of radio waves. 
In an experiment, the radiation energy can be determined by interpolating and integrating the measured energy fluence on the ground (cf.\ Sec.~\ref{sec:experimenterad} or \cite{AERAEnergyPRL,AERAEnergyPRD}). As soon as the air shower has emitted all radio emission at the end of the shower development, the radiation energy measured at increasing atmospheric depths remains constant because the atmosphere is essentially transparent for radio emission. In particular, the radiation energy is independent of the signal distribution on the ground that changes drastically with incoming direction of the air shower or the altitude of the observation.

The CoREAS simulation code is used to simulate air showers and their radio emission \cite{Corsika,CoREAS2013}. CoREAS is a microscopic simulation code, where each air-shower particle is tracked and its radiation is calculated from first-principles following the outcome of classical electrodynamics that radiation comes from accelerated charges. Then, the radiation from all shower particles is superimposed and results in the macroscopically observed radio pulse.
The advantage of a microscopic calculation is that the calculation does not have any free parameters. It is purely based on classical electrodynamics. Uncertainties in the simulation of the radio emission arise only from approximations made in the simulation to speed up the computation or from the simulation of the air shower itself (mainly uncertainties in hadronic interactions) or the modeling of the atmosphere. Hence, we can use the microscopic models for an quantitative prediction of the radio emission from extensive air showers.

The radiation energy originates from the radiation generated by the electromagnetic part of the air shower. Hence, the radiation energy correlates best with the energy of the electromagnetic cascade and not with the complete shower energy, which includes energy carried by neutrinos and high-energy muons that are not relevant for radio emission, the so-called \emph{invisible energy}. This is beneficial e.g. for a direct comparison with the measurement of the fluorescence technique which is also only sensitive to the electromagnetic shower component. However, if the cosmic-ray energy is the quantity of interest, a correction for the invisible energy needs to be applied which has been measured in \cite{InvisibleEnergy2013} with a systematic uncertainty of 3\%. 
In \cite{GlaserErad2016} it was shown that the radiation energy -- after correcting for the dependence on the geomagnetic field and the air-density of the emission region -- correlates with the electromagnetic shower energy with a scatter of less than 3\%. 

\section{The Engineering Radio Array of the Pierre Auger Observatory}
The Pierre Auger Observatory is the world's largest detector for high energy cosmic rays covering an area of \unit[3000]{km$^2$} \cite{Auger2014}. It is located on a vast, high plain in western Argentina known as the Pampa Amarilla (yellow prairie) East of the Andes near the town Malarg\"ue. The area is generally flat with an average height of \unit[1400]{m a.s.l.} that corresponds to an atmospheric overburden of \SI{875}{g/cm^2}.

The observatory is designed as a hybrid detector with two baseline components, a large surface detector array (SD) and a fluorescence detector (FD). The SD comprises 1660 water Cherenkov particle-detector stations distributed on a hexagonal grid with a spacing of \unit[1500]{m} which is overlooked by 24 air fluorescence telescopes positioned at four different sites around the SD array. The SD has a duty cycle of 100\% whereas the detection of fluorescence light is restricted to moonless nights with good weather conditions resulting in a duty cycle of $\sim$15\% at the present time \cite{Auger2014}. 
This hybrid design is extended by three high elevation fluorescence telescopes (HEAT) that overlook an infill array consisting of 61 stations with a smaller spacing of \unit[750]{m} covering an area of \unit[23.5]{km$^2$} in the western part of the array. This extension lowers the full efficiency threshold of the surface detector from \unit[3]{EeV} (=\SI{e18.5}{eV}) by one order of magnitude to \unit[0.3]{EeV} (=\SI{e17.5}{eV}) \cite{AUGERICRC2011_Maris}. 

\begin{figure}[t]
\centering
\includegraphics[width=0.9\textwidth,clip]{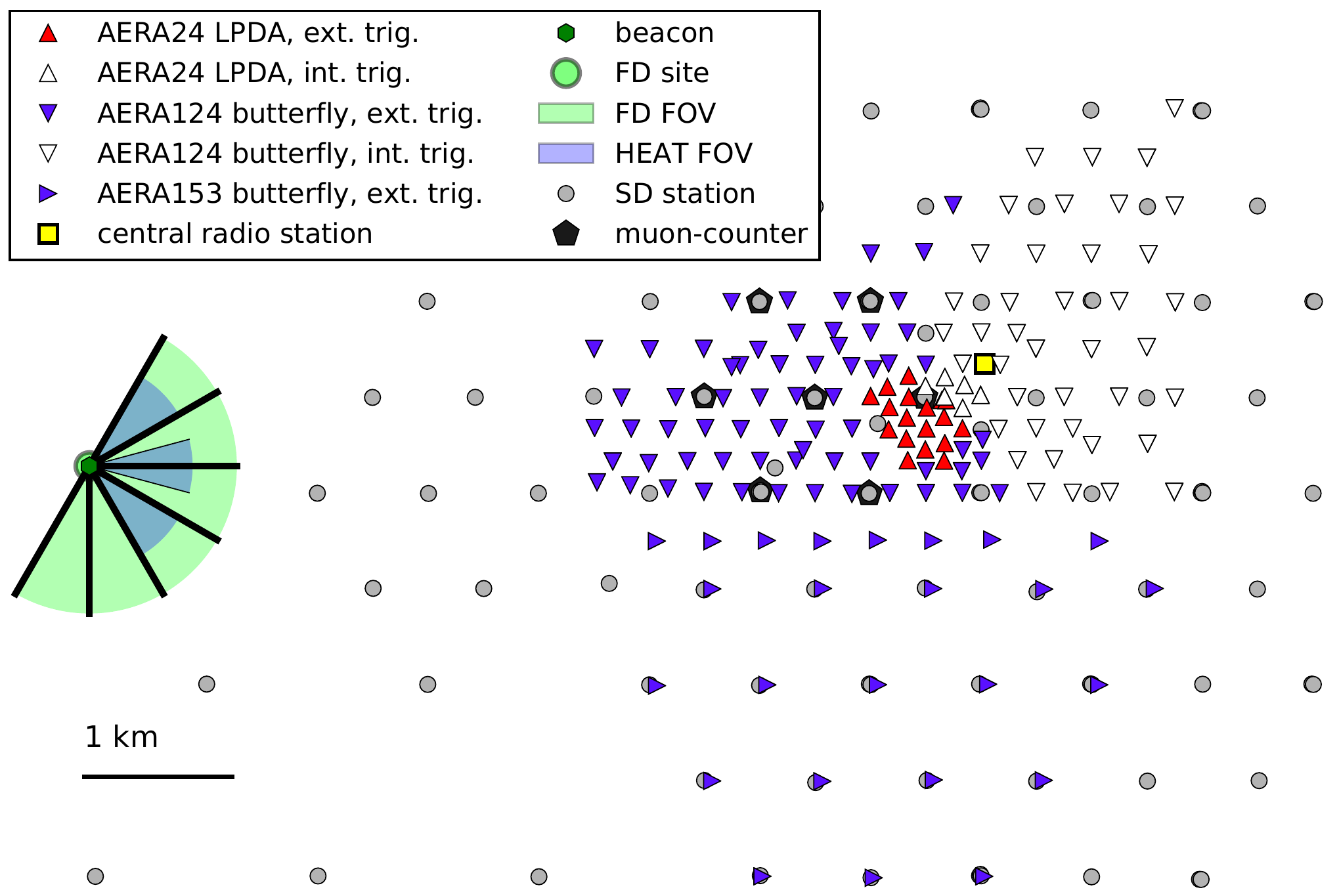}
\caption{Map of AERA within the Pierre Auger Observatory. The radio detector stations (triangles) equipped with different antennas and digitizing hardware are surrounded by surface detector stations (gray filled circles) and underground muon counters \cite{AMIGA2016} (black pentagons). AERA is in the field of view of the Coihueco and HEAT fluorescence telescopes. The orientation of the triangles indicate the three stages of expansion, upturned triangles the first (AERA24), down-turned triangles the second (AERA124) and right-turned triangles the third stage of expansion (AERA153). The color of the triangles indicate the version of the electronics. Figure adapted from \protect \cite{ICRC2015JSchulz}.}
\label{fig:map_AERA} 
\end{figure}

The Auger Engineering Radio Array (AERA) is the radio extension of the Pierre Auger Observatory, located in its western part within the \SI{750}{m} spaced surface detector array and in the field of view of the Coihueco and HEAT fluorescence telescopes. In its current stage of expansion it covers an area of \SI{17}{km^2} and consists of more than 150 autonomously operating radio detector stations (RDS) sensitive to the frequency range of \unit[30 to 80]{MHz}. A map of AERA is presented in Fig.~\ref{fig:map_AERA}. AERA measures air showers with energies above \SI{e17}{eV} which coincides with the energy thresholds of SD and HEAT and thus allows for a coincident detection of air showers.  

All radio stations operate autonomously. They are solar powered and equipped with a battery to run \SI{24}{h} each day. Accurate timing is achieved with GPS receivers at each station and an additional method to calibrate the relative timing of the radio stations to better than \SI{2}{ns} using a reference beacon signal \cite{BeaconAirplane2016}. The inner 24 stations with a spacing of \SI{144}{m} are equipped with log-periodic dipole antennas (LPDAs) \cite{AntennaPaper}. All other stations are equipped with a so-called Butterfly antenna, which is a broad-band dipole antenna of the family of bow-tie antennas \cite{AntennaPaper}. All stations operate in the frequency band from \unit[30 - 80]{MHz}.

All AERA radio stations are thoroughly calibrated. The most challenging part of the detector calibration is the absolute calibration of the antenna response to cosmic-ray radio signals. To be unaffected by near-field effects, a calibrated signal source needs to be placed at least $\SI{\sim 20}{m}$ away from the antenna at any place around the antenna. Several calibration campaigns were performed and the measurement setup was continuously improved. The first campaign used a weather balloon to lift the calibrated signal source and measured the horizontal component of antenna response with an accuracy of 12.5\% \cite{AERAEnergyPRD,AntennaPaper}. 
By now, an improved method has been developed that uses a remotely piloted drone to lift the signal source \cite{AERACalibPaper}. This setup is much less dependent on wind but has a more limited payload. 
This setup was completed with an optical camera system to accurately determine the position of the signal source during the flight \cite{ARENA2016Briechle}. Two standard digital cameras are placed at to orthogonal axes at $\sim$\SI{100}{m} distance from the antenna. From the photos taken every few seconds, the position of the signal source can be reconstructed. This information is combined with the position information from the GPS and the barometer of the drone to achieve a small statistical as well as systematic uncertainty in the position of the signal source. 
Using this new method an accuracy of 9\% was achieved for the LPDA antennas with respect to the measured energy fluence of the radio signal \cite{AERACalibPaper}.

\section{Measurement of the cosmic-ray energy}
\label{sec:experimenterad}

To measure the cosmic-ray energy, we introduce a general approach with a direct physical interpretation \cite{AERAEnergyPRL,AERAEnergyPRD}. At each observer position we calculate the energy deposit per area of the cosmic-ray radio pulse and by integrating the two-dimensional lateral distribution function (LDF) over the area we obtain the total amount of energy that is transferred from the primary cosmic ray into radio emission during the air-shower development, the \emph{radiation energy}. In Fig.~\ref{fig:LDF} left, we present the radio signal distribution of a measured cosmic-ray and the corresponding LDF fit. This approach is independent of the shape of the signal distribution because energy, i.e., the integral over the signal distribution, is conserved.

In Fig.~\ref{fig:LDF} right, we present the relation between the cosmic-ray energy and the radiation energy for primaries of energy in the EeV (= \unit[10\textsuperscript{18}]{eV}) range. To obtain this relation we use data of the LPDA radio stations of AERA and take advantage of the possibility to cross-calibrate these measurements with the well-understood data of the Pierre Auger Observatory. As the strength of the dominant geomagnetic emission process scales with the angle $\upalpha$ between shower axis and geomagnetic field, the radiation energy $E_\mathrm{30 - 80 MHz}^\mathrm{Auger}$ is divided by $\sin^2 \upalpha$.  We observe that the corrected radiation energy scales quadratically with the cosmic-ray energy as expected for coherent emission. The fit of a calibration function yields a slope of \num{1.98+-0.04}. 

From the fit of the calibration curve, we obtain the relation between radiation energy and cosmic-ray energy. We normalize the relation to the geomagnetic field strength of the Auger site to make our measurement applicable at any location on Earth except for detectors at high altitudes where a significant fraction of the air shower is clipped away at the ground. The measured relation is
\begin{equation}
 E_\mathrm{\unit[30 - 80]{MHz}} = \unit[(15.8 \pm 0.7 (\mathrm{stat}) \pm 6.7 (\mathrm{sys}))]{MeV}  
 \times \left(\sin\alpha \,\frac{E_\mathrm{CR}}{\unit[10^{18}]{eV}} \, \frac{B_\mathrm{Earth}}{\unit[0.24]{G}} \right)^2 ,
\end{equation}
where $E_\mathrm{CR}$ is the cosmic-ray energy, $B_\mathrm{Earth}$ denotes the local magnetic-field strength and \unit[0.24]{G} is the magnetic-field strength at the Auger site. The systematic uncertainty quoted here is the combined uncertainty of the radiation energy (28\%) and the SD energy scale (16\% at \unit[$10^{17.5}$]{eV}). We note that the uncertainty of the cosmic-ray energy is only half of the uncertainty of $E_\mathrm{\unit[30 - 80]{MHz}}$ as the radiation energy scales quadratically with the cosmic-ray energy. 

From the scatter around the calibration curve, the energy resolution of the AERA detector can be determined. We obtain an energy resolution of 22\% for the full data set and a reduced uncertainty of 17\% for a subset with at least five stations with signal above threshold per event.

\section{Independent determination of the cosmic-ray energy}
Instead of calibrating the measured radiation energy with the measurement of the cosmic-ray energy by the baseline detectors of the Pierre Auger Observatory, we can use the theoretical prediction of the radiation energy to determine the cosmic-ray energy independently with the radio technique. This method is currently exploited at the Pierre Auger Observatory. In this section, we present the method and discuss our current estimate of the systematic uncertainties.

\begin{figure}[tb]
\centering
 \includegraphics[trim={0 1.3cm 0 2.2cm},clip,width=1\textwidth]{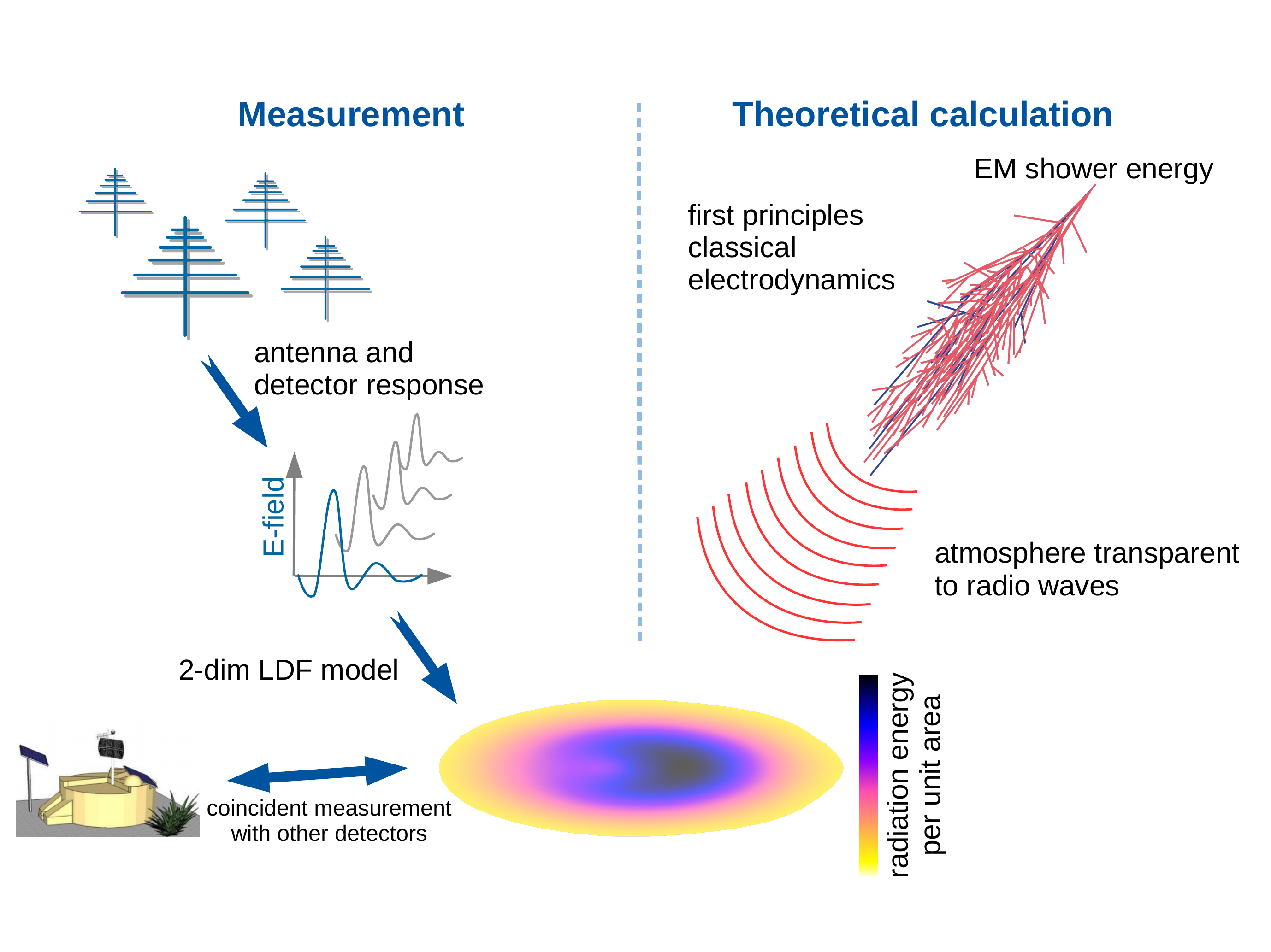}
 \caption{Illustration of the determination of the energy scale from first-principles calculations using the radio technique. The method is divided into an experimental part (the measurement of the radiation energy) and a theoretical part (the prediction of the radiation energy).}
 \label{fig:energyscale_sketch}
\end{figure}

The full procedure to determine the energy scale with a radio detector is depicted in Fig.~\ref{fig:energyscale_sketch}. The procedure can be subdivided into an experimental part, the measurement of the radiation energy (see previous section), and a theoretical part, where the radiation energy is calculated from first principles. 
On the theoretical side which was briefly discussed in Sec.~\ref{sec:theoryerad}, the radiation energy is calculated via first principles from the electromagnetic air-shower component using classical electrodynamics (see \cite{GlaserErad2016} for a detailed discussion). As the atmosphere is transparent to radio waves, the predicted radiation energy can be directly compared to the measurement. 

The systematic uncertainties contributing to the energy measurement can be subdivided into different categories. They are summarized in Tab.~\ref{tab:energyscale_uncertainties} and are briefly discussed in the following (see \cite{PhDGlaser} for a more comprehensive discussion). All uncertainties are given with respect to the cosmic-ray energy \footnote{As the radiation energy scales quadratically with the cosmic-ray energy, an uncertainty of $X\%$ on the radiation energy corresponds to an uncertainty of $(0.5 \times X)\%$ on the cosmic-ray energy.}. 

\paragraph{Experimental uncertainties} 
The uncertainties of the measurement of the radiation energy that were presented in Fig.~\ref{fig:LDF} right and in \cite{AERAEnergyPRL,AERAEnergyPRD} amount to 14\% and are dominated by the uncertainty of the antenna response of 12.5\% and the calibration of the signal chain of 6\%. Also the systematic uncertainty of the LDF model contributes with 2.5\%.

By now, an improved detector calibration was performed that lowers the uncertainty of the antenna response to 9\% \cite{AERACalibPaper}. 
Also the uncertainty of the signal chain was reduced to below 1\% in a new measurement. Hence, the experimental uncertainties of the measurement of the radiation energy could be reduced to below 10\%.

\newcolumntype{.}{D{.}{.}{-1}}
\newcommand{\specialcell}[2][c]{%
  \begin{tabular}[#1]{@{}c@{}}#2\end{tabular}}

\newcommand\tw{0.5cm}
\newcommand\tww{0.2cm}
\begin{table}[tbp]
\centering
\caption{\label{tab:energyscale_uncertainties}Estimate of the systematic uncertainties of the cosmic-ray energy measurement with the AERA detector. See text for details.} 
 \begin{tabularx}{0.6\textwidth}{Xc}
 \hline \hline 
  \rule{0pt}{3ex}\textbf{source of uncertainty} &  \\ \hline
  \rule{0pt}{4ex}\textbf{experimental uncertainties} &  \textbf{9.4\%}   \\
  \hspace*{\tww} antenna response pattern \cite{AERACalibPaper} & \hspace*{\tw}9\% \\
  \hspace*{\tww} analog signal chain  & \hspace*{\tw}$<$1\%  \\
  \hspace*{\tww} LDF model & \hspace*{\tw} $<$2.5\%  \\

  \rule{0pt}{4ex}\textbf{theoretical uncertainties}  & \textbf{2\%}   \\

\rule{0pt}{4ex}\textbf{environmental uncertainties} & \textbf{1.6\%}   \\
  \hspace*{\tww} atmosphere  & \hspace*{\tw}1.25\% \\
  \hspace*{\tww} ground conditions \cite{AERACalibPaper}  & \hspace*{\tw}1\% \\
  
\rule{0pt}{4ex}\textbf{invisible energy correction} \cite{InvisibleEnergy2013} & \textbf{3.0\%}  \\

\rule{0pt}{3ex}\textbf{total absolute scale uncertainty}  & \textbf{10.2\%} \\ \hline \hline
 \end{tabularx}
\end{table}

\paragraph{Theoretical uncertainties}
The calculation of the radio emission from the electromagnetic shower particles itself has no systematic uncertainties as the calculation is purely based on classical electrodynamics which does not have any free parameters. However, the modeling of the electromagnetic air-shower component is subject to uncertainties. Also, approximations made in the simulation code to speed up the calculations result in additional uncertainties. 

The systematic uncertainty in modeling the air-shower development is estimated by using different (high- and low-energy) hadronic interaction models. Exchanging QGSJetII-04 with EPOS-LHC and FLUKA with UrQMD results in a difference in the predicted radiation energy of 0.18\% each which is compatible within the statistical uncertainty of the comparison \cite{GlaserErad2016}. The influence of hadronic interaction models is small because we correlate the radiation energy directly with the electromagnetic shower energy. We use the differences in radiation energy as an estimate of the systematic uncertainty. We add the two differences of the low- and high-energy hadronic interaction models in quadrature resulting in an uncertainty with respect to the cosmic-ray energy of below 0.13\%.

The impact of thinning and lower energy thresholds of the shower particles was studied and contributes with 0.15\% and 0.5\% to the systematic uncertainty \cite{GlaserErad2016}.

An independent cross-check of a correct implementation of the underlying physics in the CoREAS simulation code \cite{CoREAS2013} was performed by a detailed comparison with the completely independent ZHAireS simulation code \cite{ZHAires2012}. The difference in the prediction of the radiation energy from the two codes is 3.3\% and accordingly 1.7\% with respect to the cosmic-ray energy \cite{MAMarvin}. We use this difference as an additional contribution to the systematic uncertainty. 

In total, we estimate the theoretical uncertainties to contribute with $\sim$2\% to the systematic uncertainty. In this estimation we studied all known influences on the radiation energy. However, we can not exclude the existence of unknown additional effects that have larger influences than the estimated 2\%, although it seems unlikely as the measured form of the radio pulse as well as the signal distribution on ground is reproduced by the CoREAS simulations with high precision (e.g. \cite{LOFARNature2016}) and so far no features of the radio emission were observed that are incompatible with the Monte-Carlo prediction. 

The calculation of the radio emission can also be probed by lab experiments. The measurement presented in \cite{SlacT510} is compatible with the Monte-Carlo prediction. However, the systematic uncertainty of the measurement is in the order of 30\% and accordingly too large to be used here as a relevant benchmark.

\paragraph{Environmental uncertainties}
Changing atmospheric conditions, i.e., changing density profiles and varying refractivity, result in a scatter of 1.25\% during the course of the year. We use this scatter as an estimate of the systematic uncertainty as we use an average state of the atmosphere in the calculation. To further reduce this uncertainty, separate simulations for monthly averages of the atmospheric profiles can be performed. Even separate simulations for each measured event are possible where the respective atmospheric conditions at the time of the event are obtained from GDAS data \cite{GDASAuger2012}.

Changing ground conditions impacts the reflectivity of the ground and thereby the antenna response. In case of the LPDAs, the antenna sensitivity towards the ground is small. A simulation of the antenna response pattern with realistic variations of ground parameters results in a change of the antenna response of $\sim$1\% \cite{AERACalibPaper,PhDKrause}. This uncertainty could be further reduced with a continuous monitoring of the ground conditions. 

In total, the environmental uncertainties contribute with 1.6\% to the systematic uncertainty.

\paragraph{Invisible energy correction}
To obtain the cosmic-ray energy from the electromagnetic shower energy, the invisible energy needs to be taken into account. This can be done using the parametrization that was obtained from a measurement and has a systematic uncertainty of 3\% at \SI{e18}{eV} \cite{InvisibleEnergy2013}. 

\section{Future potential of the radio technique}
The accurate energy measurement of a radio detector can be used to improve the energy calibration of cosmic-ray observatories. Established detection methods such as fluorescence telescopes or surface detector arrays can be compared with the radio technique by coincident measurement of air showers. In particular, the combination of the radio and the fluorescence technique offers great potential: Both methods are directly sensitive to the electromagnetic air-shower component and the systematic uncertainties of the two techniques are mostly independent. Hence, systematic uncertainties could be further reduced by combining the two independent approaches for the absolute energy calibration.

\section{Conclusions}
Ultra-high energy cosmic rays can be measured by broadband MHz radio emission from air showers. This technique is especially useful to accurately determine the cosmic-ray energy. Its main advantages are that the atmosphere is transparent to radio waves and that the radio emission can be calculated by first-principles from the air shower development. Hence, the systematic uncertainty in the energy measurement is dominated by the detector calibration which could be reduced to $\sim$9\% in a recent calibration campaign of the AERA detector. The currently known systematic uncertainties arising from the theoretical calculation are estimated to $\sim$2\% by studying the effect of all known influences. In total, we estimate the total currently achievable systematic uncertainty with the AERA detector to 10\%.

\section*{References}

\end{document}